\documentclass[twocolumn,twoside,slac_two]{revtex4}
\usepackage{graphicx}
\usepackage{fancyhdr}
\def\beq{\begin{equation}}
\def\eeq{\end{equation}}
\def\beqa{\begin{eqnarray}}
\def\eeqa{\end{eqnarray}}

\begin{document}

\title{Two-loop soft anomalous dimensions with massive and massless quarks}

%

\author{Nikolaos Kidonakis}
\affiliation{Kennesaw State University, Kennesaw, GA 30144, USA}
\begin{abstract}
I present results for two-loop soft anomalous dimensions, which
are  derived from dimensionally regularized diagrams with eikonal 
quark lines and control soft-gluon emission in hard-scattering processes.
Detailed results for the UV poles of the eikonal integrals are 
shown for massive quarks, and the massless limit is also taken. 
The construction of soft anomalous dimensions at two-loops
allows soft-gluon resummations at NNLL accuracy.
\end{abstract}

\maketitle

\thispagestyle{fancy}


\section{Introduction}

Soft-gluon resummation formalisms for top quark production and other 
hard scattering cross sections beyond leading logarithms are formulated in 
terms of exponentials of soft anomalous dimensions that control noncollinear
soft-gluon emission (see e.g. \cite{NKrev}). 
The calculations of these anomalous dimensions \cite{NKGS,NKprl} are
performed using the eikonal approximation, which is valid for descibing the 
emission of soft gluons from partons in the hard scattering.
The approximation leads to a simplified form of the Feynman rules by
removing the Dirac matrices from the calculation.
When the gluon momentum, $k$, goes to zero,
the Feynman rule for the quark-gluon vertex reduces to
$g_s T_F^c v^{\mu} / v\cdot k$ 
with $v$ a dimensionless velocity vector.

Here we calculate diagrams and derive the soft anomalous dimension 
for top quark pair production via 
$e^+ e^-\rightarrow t {\bar t}$ through two loops \cite{NKprl}. 
We also discuss the massless limit and extensions to other processes.

Writing the soft anomalous dimension $\Gamma_S$ as a series in $\alpha_s$ 
\beq
\Gamma_S=\frac{\alpha_s}{\pi} \Gamma_S^{(1)}
+\left(\frac{\alpha_s}{\pi}\right)^2 \Gamma_S^{(2)}+\cdots
\eeq
we calculate the one- and two-loop expressions for $\Gamma_S$.
We use the Feynman gauge and we employ dimensional
regularization with $n=4-\epsilon$ dimensions.
The soft anomalous dimension can be determined from the coefficients  
of the ultraviolet (UV) poles of the eikonal diagrams. 

We do not include the color factors in the individual diagrams below, but 
take them into account when assembling all the pieces together in the 
last section. The Appendix lists some of the integrals calculated for the 
evaluation of the diagrams.

\section{One-loop diagrams}

\begin{figure}
\begin{center}
\includegraphics[width=8cm]{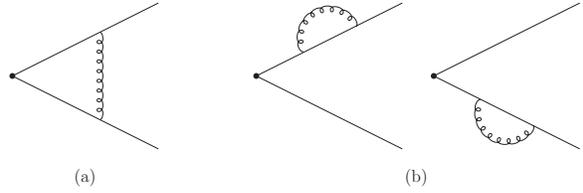}
\caption{One-loop diagrams with heavy-quark eikonal lines.}
\end{center}
\label{1loop}
\end{figure}

In this section we calculate the one-loop diagrams in Fig. 1.
We define $\beta=\sqrt{1-4m^2/s}$ with $m$ the heavy quark mass and 
$s$ the squared c.m. energy.
We label the two heavy quark lines by $i$ and $j$. 
Using the relation $v=p \, {\sqrt{2/s}}$, with $p$ the momentum, we have  
$v_i^2=v_j^2=(1-\beta^2)/2$ and $v_i \cdot v_j=(1+\beta^2)/2$.

We begin with the integral $I_{1a}$ for the 
diagram in Fig. 1(a) given by
\beq
I_{1a} = g_s^2 \int\frac{d^n k}{(2\pi)^n} \frac{(-i)g_{\mu \nu}}{k^2} 
\frac{v_i^{\mu}}{v_i\cdot k} \, \frac{(-v_j^{\nu})}{(-v_j\cdot k)} \, .
\eeq
Using Feynman parameterization and after several manipulations and 
separation of UV and infrared (IR) singularities, 
we find through Eq. (\ref{A1}) the UV poles
\beq
I_{1a}=\frac{\alpha_s}{\pi} \frac{(1+\beta^2)}{2\, \beta}
\frac{1}{\epsilon} \ln\left(\frac{1-\beta}{1+\beta}\right) \, . 
\label{I1aUV}
\eeq

Next we calculate the one-loop self-energy diagrams in Fig. 1(b) given by 
\beq
I_{1b}=g_s^2 \int\frac{d^n k}{(2\pi)^n} \frac{(-i)g_{\mu \nu}}{k^2} 
\frac{v_i^{\mu}}{v_i\cdot (k'-k)} \, \frac{v_i^{\nu}}{v_i\cdot k'} \, . 
\eeq
Here we have introduced the regulator $k'$, i.e. the external quark momentum is 
$p_i+k'$, so as to use the eikonal rules. 
We then expand the above expression around $v_i\cdot k'=0$ at constant 
$\epsilon$. The expansion gives an irrelevant $1/v_i \cdot k'$ term, 
a constant term, 
and linear and higher-order terms in $v_i \cdot k'$ which vanish when 
setting $k'=0$. The remaining term is then
\beq
I_{1b}=\frac{i \, g_s^2 \, v_i^2}{(2\pi)^{n}}
\int \frac{d^n k}{k^2 \, (v_i\cdot k)^2} \, . 
\eeq
We isolate the UV poles and using Eq. (\ref{A2}) we find 
\beq
I_{1b}=\frac{\alpha_s}{\pi} \frac{1}{\epsilon} \, .
\label{I1bUV}
\eeq

The one-loop soft anomalous dimension is then read off the coefficient 
of the UV poles of the one-loop diagrams through
\beq
C_F[I_{1a}+I_{1b}]=-\frac{\alpha_s}{\pi} \frac{\Gamma_S^{(1)}}{\epsilon}
\eeq
which gives 
\beq
\Gamma_S^{(1)}=C_F \left[-\frac{(1+\beta^2)}{2 \, \beta} 
\ln\left(\frac{1-\beta}{1+\beta}\right) -1\right]
\label{GammaS1}
\eeq
with $C_F=(N_c^2-1)/(2 N_c)$ the color factor. 

When expressed in terms of the cusp angle 
$\gamma=\cosh^{-1}\left(v_i \cdot v_j/\sqrt{v_i^2 v_j^2}\right)
=\ln[(1+\beta)/(1-\beta)]$ this becomes 
$\Gamma_S^{(1)}=C_F (\gamma \coth\gamma-1)$ and 
is also known as a cusp anomalous dimension \cite{IKR,KR}. 

\section{Two-loop vertex diagrams}

\begin{figure}
\begin{center}
\includegraphics[width=8cm]{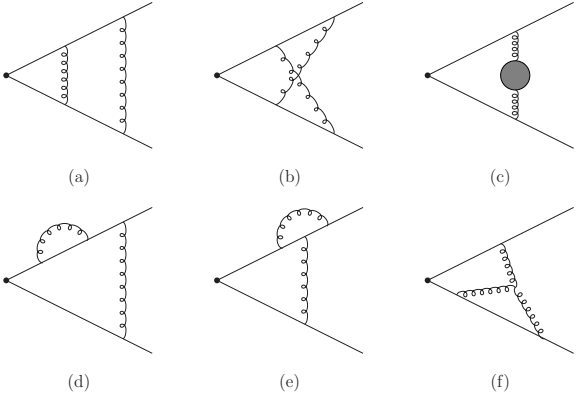}
\caption{Two-loop vertex diagrams with heavy-quark eikonal lines.}
\end{center}
\label{2vloop}
\end{figure}

The two-loop diagrams can be separated into two groups, shown in Fig. 2 and 3.
The calculations are 
challenging due to the presence of a heavy quark mass and they 
involve multiple complicated integrals and delicate separations 
of infrared and ultraviolet poles, which by construction of the 
soft function are opposites of each other \cite{NKGS}. 
The analytical structure of the results involves logarithms and polylogarithms.

In this section we calculate the two-loop vertex diagrams of Fig. 2.  
We also have to include the one-loop counterterms to these diagrams.
We will denote by $I'_2$ the diagrams in Fig. 2 and by $I^{c.t.}_2$ their 
counterterms. The total contribution will then be $I_2=I'_2+I^{c.t.}_2$.

We find it convenient to define 
\beqa
M_{\beta}&=&\left(4 \ln 2+\ln \pi-\gamma_E-i\pi\right)
\ln\left(\frac{1-\beta}{1+\beta}\right) 
\nonumber \\ &&  
{}+\frac{1}{2} \ln^2(1+\beta)-\frac{1}{2} \ln^2(1-\beta)
\nonumber \\ && 
{}-{\rm Li}_2\left(\frac{1+\beta}{2}\right)
+{\rm Li}_2\left(\frac{1-\beta}{2}\right) \, . 
\eeqa

\subsection{$I_{2a}$ and $I_{2b}$}

In this section we calculate the integrals $I_{2a}$ and $I_{2b}$ 
for the two-loop diagrams in Figs. 2(a) and 2(b). 
Note that there is no counterterm for diagram 2(b), so $I_{2b}=I'_{2b}$.

Diagram 2(b) is given by
\beqa
&& \hspace{-12mm}
I_{2b}=g_s^4 \int\frac{d^n k_1}{(2\pi)^n}\frac{d^n k_2}{(2\pi)^n}
\frac{(-i)g_{\mu\nu}}{k_1^2} \frac{(-i)g_{\rho\sigma}}{k_2^2}  
\nonumber \\ && \hspace{-9mm} \times 
\frac{v_i^{\mu}}{v_i\cdot k_1} \frac{v_i^{\rho}}{v_i\cdot (k_1+k_2)}
\frac{(-v_j^{\nu})}{-v_j\cdot (k_1+k_2)} \frac{(-v_j^{\sigma})}{-v_j\cdot k_2}.
\eeqa

Now, we begin with the $k_2$ integral and use Feynman parameterization. 
Performing first the $k_2$ integration and then an integration over one of 
the parameters, we have
\beqa
&& \hspace{-3mm}
I_{2b}= -i \frac{\alpha_s^2}{\pi^2} \, 2^{-4+\epsilon} \, 
\pi^{-2+3\frac{\epsilon}{2}} \, 
\Gamma\left(1-\frac{\epsilon}{2}\right) \, \Gamma(1+\epsilon)
\nonumber \\ && 
\times (1+\beta^2)^2 \int_0^1 dz \int_0^1 dy \, (1-y)^{-\epsilon}
\nonumber \\ &&  \times 
\left[2\beta^2(1-y)^2 z^2
-2\beta^2(1-y)z-\frac{(1-\beta^2)}{2}\right]^{-1+\frac{\epsilon}{2}} 
\nonumber \\ && 
\times \int \frac{d^n k_1}{k_1^2 \, v_i \cdot k_1 \, 
\left[\left((v_i-v_j)z+v_j\right)\cdot k_1\right]^{1+\epsilon}} \, .
\label{I2b}
\eeqa
Next we proceed with the $k_1$ integral
\beqa
&& \hspace{-5mm}\int \frac{d^nk_1}{k_1^2 \, v_i \cdot k_1 \, 
\left[\left((v_i-v_j)z+v_j\right)\cdot k_1\right]^{1+\epsilon}}=
-i (-1)^{-3\frac{\epsilon}{2}} 
\nonumber \\ && \hspace{-3mm} \times 
2^{2+3\epsilon} \pi^{2-\frac{\epsilon}{2}} 
\frac{\Gamma\left(1+\frac{3\epsilon}{2}\right)}{\Gamma(1+\epsilon)} 
\int_0^1 dx_1 x_1^{-1+2\epsilon} (1-x_1)^{-1-2\epsilon}
\nonumber \\ && \hspace{-3mm}
\times \int_0^1 dx_2 (1-x_2)^{\epsilon} 
\left\{\left[x_2 v_i \right. \right.
\nonumber \\ && \hspace{10mm} \left. \left.
+(1-x_2)\left((v_i-v_j)z+v_j\right)\right]^2
\right\}^{-1-3\frac{\epsilon}{2}}.
\eeqa
The integral over $x_1$ contains both UV and IR singularities. We isolate 
the UV singularities via
\beqa
&& \hspace{-11mm}\int_0^1 dx_1 \, x_1^{-1+2\epsilon} (1-x)^{-1-2\epsilon}
=\int_0^1 dx_1 \, x_1^{-1+2\epsilon}
\nonumber \\ && \hspace{-10mm} 
{}+\int_0^1 dx_1 \, x_1^{-1+2\epsilon}
\left[(1-x)^{-1-2\epsilon}-1\right]
=\frac{1}{2\epsilon}+{\rm IR}.
\eeqa
We set $\epsilon=0$ in the integral over $x_2$ since it is UV finite. 
We thus find the UV poles of the $k_1$ integral 
\beqa
&& \hspace{-10mm} \int \frac{d^nk_1}{k_1^2 \, v_i \cdot k_1 \, 
\left[\left((v_i-v_j)z+v_j\right)\cdot k_1\right]^{1+\epsilon}}
=-i \, 2\pi^2 \frac{1}{\epsilon} 
\nonumber \\ && \hspace{-8mm} \times 
\frac{1}{\beta (1-z)}
\left[\tanh^{-1}(\beta(1-2z))-\tanh^{-1}(-\beta)\right] \, .
\eeqa
Using this result in the expression for $I_{2b}$, Eq. (\ref{I2b}), 
and performing the $y$ and $z$ integrals, we find 
\beqa
&& \hspace{-6mm}
I_{2b}=\frac{\alpha_s^2}{\pi^2} \frac{(1+\beta^2)^2}{8 \, \beta^2} 
\frac{1}{\epsilon} \left\{-\frac{1}{3}\ln^3\left(\frac{1-\beta}{1+\beta}\right)
\right.
\nonumber \\ && \hspace{15mm} 
{}-\ln\left(\frac{1-\beta}{1+\beta}\right)
\left[{\rm Li}_2\left(\frac{(1-\beta)^2}{(1+\beta)^2}\right)
+\zeta_2\right]
\nonumber \\ && \hspace{15mm} \left. 
{}+{\rm Li}_3\left(\frac{(1-\beta)^2}{(1+\beta)^2}\right)-\zeta_3 \right\}.
\label{I2bUV}
\eeqa

Diagrams 2(a) and 2(b) are related through the equation 
$I'_{2a}+I_{2b}=\frac{1}{2} I_{1a}^2$ (where we need to keep UV poles 
and constant terms in $I_{1a}$) from which we can calculate $I'_{2a}$.
The one-loop counterterm to $I_{2a}$ is 
\beqa
&& \hspace{-15mm}
I_{2a}^{c.t.}=\frac{\alpha_s^2}{\pi^2} \frac{(1+\beta^2)^2}{8 \, \beta^2} 
\left\{
-\frac{2}{\epsilon^2} \ln^2\left(\frac{1-\beta}{1+\beta}\right) \right.
\nonumber \\ && \hspace{20mm}
-\frac{1}{\epsilon} M(\beta) \ln\left(\frac{1-\beta}{1+\beta}\right) \, .
\eeqa
Then we find the simple relation 
\beq
I_{2a}+I_{2b}=\frac{\alpha_s^2}{\pi^2} \frac{(1+\beta^2)^2}{8 \, \beta^2} 
\frac{(-1)}{\epsilon^2} 
\ln^2\left(\frac{1-\beta}{1+\beta}\right) \, .
\eeq

\subsection{$I_{2c}$}

In this section we calculate the integrals for the three diagrams 
represented by Fig. 2(c). The blob represents a quark, gluon, or ghost loop.
Note that an additional diagram with a four-gluon vertex vanishes.

The quark-loop diagram is given by
\beqa
&& \hspace{-10mm} 
I'_{2cq}=(-1) n_f g_s^4 \int\frac{d^n k}{(2\pi)^n}\frac{d^n l}{(2\pi)^n}
\frac{v_i^{\mu}}{v_i\cdot k} \frac{(-v_j^{\rho})}{(-v_j\cdot k)}
\nonumber \\ && \hspace{-8mm} \times
\frac{(-i)g_{\mu\nu}}{k^2} \frac{(-i)g_{\rho\sigma}}{k^2} 
{\rm Tr} \left[-i \gamma^{\nu}
\frac{i l\!\!/}{l^2} (-i) \gamma^{\sigma} i \frac{(l\!\!/-k\!\!/)}
{(l-k)^2}\right].
\label{I2cq}
\eeqa
After several steps, and using Eq. (\ref{A3}), the final result for 
the UV poles of $I'_{2cq}$ is
\beqa
&& \hspace{-12mm}
I'_{2cq}=\frac{\alpha_s^2}{\pi^2} n_f \frac{(1+\beta^2)}{6\, \beta}
\left\{-\frac{1}{\epsilon^2} \ln\left(\frac{1-\beta}{1+\beta}\right) \right.
\nonumber \\ && \hspace{17mm} \left.
{}-\frac{1}{\epsilon}\left[\frac{5}{6}\ln\left(\frac{1-\beta}{1+\beta}\right)
+M_{\beta}\right]\right\}
\label{I2cqUV}
\eeqa
with $n_f$ the number of light quark flavors.
The one-loop counterterm is
\beq
I_{2cq}^{c.t.}=\frac{\alpha_s^2}{\pi^2} n_f \frac{(1+\beta^2)}{6\, \beta}
\left\{\frac{2}{\epsilon^2} \ln\left(\frac{1-\beta}{1+\beta}\right) 
+\frac{1}{\epsilon} M_{\beta}\right\} \, .
\label{I2cqct}
\eeq
Then the sum $I_{2cq}=I'_{2cq}+I_{2cq}^{c.t.}$ gives
\beq
I_{2cq}=\frac{\alpha_s^2}{\pi^2} n_f \frac{(1+\beta^2)}{6 \, \beta}
\left[\frac{1}{\epsilon^2}-\frac{5}{6 \, \epsilon}\right] 
\ln\left(\frac{1-\beta}{1+\beta}\right).
\eeq

Next we calculate the integral for the gluon-loop diagram 
given by
\beqa
&& \hspace{-4mm} 
I'_{2cgl}=\frac{1}{2} g_s^4 \int\frac{d^n k}{(2\pi)^n}\frac{d^n l}{(2\pi)^n}
\frac{v_i^{\mu}}{v_i\cdot k} \frac{(-v_j^{\nu})}{(-v_j\cdot k)}
\nonumber \\ &&  \times 
\frac{(-i)g_{\mu\mu'}}{k^2} \frac{(-i)g_{\rho\rho'}}{l^2} 
\frac{(-i)g_{\sigma\sigma'}}{(k-l)^2} \frac{(-i)g_{\nu\nu'}}{k^2}
\nonumber \\ &&  \times
\left[g^{\mu' \rho} (k+l)^{\sigma}+g^{\rho \sigma} (k-2l)^{\mu'}
+g^{\sigma \mu'} (-2k+l)^{\rho}\right]
\nonumber \\ &&  \times
\left[g^{\rho' \nu'} (l+k)^{\sigma'}+g^{\nu' \sigma'} (-2k+l)^{\rho'}\right.
\nonumber \\ && \hspace{30mm} \left. 
{}+g^{\sigma' \rho'} (k-2l)^{\nu'}\right]. 
\eeqa
Using Eq. (\ref{A3}) we find for the UV poles of $I'_{2cgl}$,
\beqa
&& \hspace{-12mm}
I'_{2cgl}=\frac{\alpha_s^2}{\pi^2} \frac{19}{96}  
\frac{(1+\beta^2)}{\beta}
\left\{-\frac{1}{\epsilon^2} \ln\left(\frac{1-\beta}{1+\beta}\right)\right.
\nonumber \\ && \hspace{15mm} \left.
-\frac{1}{\epsilon}\left[\frac{58}{57}\ln\left(\frac{1-\beta}{1+\beta}\right)
+M_{\beta}\right] \right\}.
\eeqa
The one-loop counterterm is 
\beq
I_{2cgl}^{c.t.}=\frac{\alpha_s^2}{\pi^2} \frac{19}{96}  
\frac{(1+\beta^2)}{\beta}
\left\{\frac{2}{\epsilon^2} \ln\left(\frac{1-\beta}{1+\beta}\right)
+\frac{1}{\epsilon} M_{\beta}\right\}.
\eeq
Then
\beq
I_{2cgl}=\frac{\alpha_s^2}{\pi^2} \frac{19}{96}  
\frac{(1+\beta^2)}{\beta}
\left\{\frac{1}{\epsilon^2}-\frac{58}{57\, \epsilon}\right\}
\ln\left(\frac{1-\beta}{1+\beta}\right) .
\label{I2cgl}
\eeq

Last we calculate the integral for the ghost-loop diagram 
given by
\beqa
&& \hspace{-10mm}
I'_{2cgh}=(-1) g_s^4 \int\frac{d^n k}{(2\pi)^n}\frac{d^n l}{(2\pi)^n}
\frac{v_i^{\mu}}{v_i\cdot k} \frac{(-v_j^{\rho})}{(-v_j\cdot k)}
\nonumber \\ && \hspace{3mm} \times 
\frac{i}{l^2} l^{\nu} \frac{i}{(l-k)^2} (l-k)^{\sigma}
\frac{(-i)g_{\mu\nu}}{k^2} \frac{(-i)g_{\rho\sigma}}{k^2}. 
\eeqa
Using Eq. (\ref{A3}), we find
\beqa
&& \hspace{-12mm}
I'_{2cgh}= \frac{\alpha_s^2}{\pi^2} 
\frac{(1+\beta^2)}{96\beta}
\left\{-\frac{1}{\epsilon^2} \ln\left(\frac{1-\beta}{1+\beta}\right) \right.
\nonumber \\ && \hspace{15mm} \left.
{}-\frac{1}{\epsilon}\left[
\frac{4}{3}\ln\left(\frac{1-\beta}{1+\beta}\right)+M_{\beta}\right]
\right\}.
\eeqa
The one-loop counterterm is
\beq
I_{2cgh}^{c.t.}= \frac{\alpha_s^2}{\pi^2} 
\frac{(1+\beta^2)}{96 \, \beta}
\left\{\frac{2}{\epsilon^2} \ln\left(\frac{1-\beta}{1+\beta}\right)
+\frac{1}{\epsilon} M_{\beta} \right\}.
\eeq
Then
\beq
I_{2cgh}=\frac{\alpha_s^2}{\pi^2} 
\frac{(1+\beta^2)}{96 \, \beta}
\left\{\frac{1}{\epsilon^2} -\frac{4}{3 \, \epsilon}\right\}
\ln\left(\frac{1-\beta}{1+\beta}\right).
\label{I2cgh}
\eeq
The sum of the gluon and ghost loops, Eqs. (\ref{I2cgl}) and (\ref{I2cgh}), 
denoted by $I_{2cg}$, is then 
\beq
I_{2cg}=\frac{\alpha_s^2}{\pi^2} \frac{5}{24} \frac{(1+\beta^2)}{\beta}  
\left[\frac{1}{\epsilon^2}-\frac{31}{30\, \epsilon}\right] 
\ln\left(\frac{1-\beta}{1+\beta}\right).
\eeq

\subsection{$I_{2d}$ and $I_{2e}$}

In this section we calculate the integral $I'_{2d}$ for the diagram in 
Fig. 2(d) given by
\beqa
I'_{2d}&=&g_s^4 \int\frac{d^n k}{(2\pi)^n}\frac{d^n l}{(2\pi)^n}
\frac{(-i)g_{\mu\nu}}{k^2} \frac{(-i)g_{\rho\sigma}}{l^2} 
\frac{v_i^{\mu}}{v_i\cdot k} 
\nonumber \\ && \quad \times 
\frac{v_i^{\sigma}}{v_i\cdot (k-l)}
\frac{v_i^{\rho}}{v_i\cdot k} \frac{(-v_j^{\nu})}{(-v_j\cdot k)} 
\eeqa
as well as the integral $I'_{2e}$ for the diagram in Fig. 2(e)
given by
\beqa
I'_{2e}&=&g_s^4 \int\frac{d^n k}{(2\pi)^n}\frac{d^n l}{(2\pi)^n}
\frac{(-i)g_{\mu\nu}}{k^2} \frac{(-i)g_{\rho\sigma}}{l^2} 
\frac{v_i^{\sigma}}{(-v_i\cdot l)} 
\nonumber \\ && \quad \times 
\frac{v_i^{\mu}}{v_i\cdot (k-l)}
\frac{v_i^{\rho}}{v_i\cdot k} \frac{(-v_j^{\nu})}{(-v_j\cdot k)} \, .
\eeqa
First we note that 
\beqa
I'_{2d}+I'_{2e}&=&\frac{g_s^4}{(2\pi)^{2n}} v_i^2 \, v_i \cdot v_j 
\int\frac{d^n k}{k^2 (v_i \cdot k)^2 v_j\cdot k}
\nonumber \\ && \quad \quad \times 
\int \frac{d^n l}{l^2 v_i \cdot l}=0
\eeqa
where the last integral is zero because it is odd in $l$.
Therefore $I'_{2d}=-I'_{2e}$ and similarly for the counterterms, 
and then $I_{2d}=-I_{2e}$.
Thus we only have to calculate $I'_{2e}$ and its counterterm.
Using Eq. (\ref{A4}) and including the counterterm 
$I_{2e}^{c.t.}=(\alpha_s^2/\pi^2) [(1+\beta^2)/(4\beta)]
[2/ \epsilon^2 \ln((1-\beta)/(1+\beta))+M_{\beta}/\epsilon]$ 
we find
\beqa
&& \hspace{-5mm}
I_{2e}=\frac{\alpha_s^2}{\pi^2} \frac{(1+\beta^2)}{4 \, \beta}
\left\{\frac{1}{\epsilon^2} \ln\left(\frac{1-\beta}{1+\beta}\right)
-\frac{1}{\epsilon} \left[\ln\left(\frac{1-\beta}{1+\beta}\right)
\right. \right.
\nonumber \\ && \quad \quad
{}+\frac{1}{2} \ln^2\left(\frac{1-\beta}{1+\beta}\right) 
-\frac{1}{2}{\rm Li}_2\left(\frac{(1- \beta)^2}{(1+\beta)^2}\right)
\nonumber \\ && \quad \quad \left. \left.
{}+\ln\left(\frac{1-\beta}{1+\beta}\right) 
\ln\left(\frac{(1+\beta)^2}{4\beta}\right)
+\frac{\zeta_2}{2}\right] \right\}. 
\eeqa

\subsection{$I_{2f}$}

In this section we calculate the three-gluon diagram in Fig. 2(f)
given by 
\beqa
&& \hspace{-3mm} I'_{2f}=g_s^4 \int\frac{d^n k_1}{(2\pi)^n} 
\frac{d^n k_2}{(2\pi)^n}
\frac{(-i)g_{\mu \mu'}}{k_1^2} \frac{(-i)g_{\nu \nu'}}{k_2^2}
\frac{(-i)g_{\rho \rho'}}{(k_1+k_2)^2} 
\nonumber \\ && \quad \quad
\times \frac{v_i^{\mu}}{v_i\cdot k_1} \, \frac{(-v_j^{\nu})}{(-v_j\cdot k_1)}  
\frac{(-v_j^{\rho})}{-v_j\cdot (k_1+k_2)} (-i) 
\nonumber \\ && \quad \quad
\times \left[g^{\mu'\nu'} (k_1-k_2)^{\rho'}
+g^{\nu'\rho'} (k_1+2k_2)^{\mu'}\right.
\nonumber \\ && \hspace{15mm} \left.
{}+g^{\rho'\mu'} (-2k_1-k_2)^{\nu'}\right]\, .
\eeqa
After many steps we find
\beqa
&& \hspace{-10mm}
I_{2f}=\frac{1}{\epsilon} \left\{
\frac{(1+\beta^2)}{12 \, \beta} 
\ln^3\left(\frac{1-\beta}{1+\beta}\right)\right. 
\nonumber \\ && 
{}-\frac{1}{4} 
\left[2 \zeta_2+\ln^2\left(\frac{1-\beta}{1+\beta}\right)\right] 
\nonumber \\ && \quad \left. \times
\left[\frac{(1+\beta^2)}{2 \, \beta} \ln\left(\frac{1-\beta}{1+\beta}\right)
+1\right]\right\}.
\eeqa

\section{Two-loop heavy-quark self-energy diagrams}

\begin{figure}
\begin{center}
\includegraphics[width=8cm]{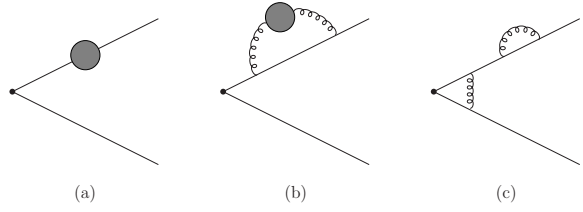}
\caption{Two-loop heavy-quark self-energy diagrams with eikonal lines.}
\end{center}
\label{2sloop}
\end{figure}

In this section we calculate the diagrams in Fig. 3 and their counterterms.
We will denote by $I'_3$ the diagrams shown in Fig. 3 and 
by $I^{c.t.}_3$ their counterterms. The total contribution is  
then $I_3=I'_3+I^{c.t.}_3$.
As for diagram 1(b) we introduce a regulator $k'$ in the quark momentum 
and then expand around $v_i\cdot k'=0$ at constant 
$\epsilon$. We also find it convient to define 
\beq
K_{\beta}= -\ln(1-\beta^2)+5 \ln 2+\ln \pi-\gamma_E-i \pi.
\eeq

\subsection{$I_{3a}$}

Here we calculate the three diagrams represented by Fig. 3(a) which are 
shown in detail in Fig. 4. 
Note that an additional graph involving a three-gluon vertex with  
all three gluons attached to the same eikonal line vanishes.

\begin{figure}
\begin{center}
\includegraphics[width=8cm]{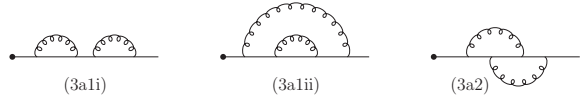}
\caption{Detail of the diagrams of Fig. 3(a).}
\end{center}
\label{3a}
\end{figure} 

We begin with the diagram labeled (3a1i). Using Eq. (\ref{A2}) we find
\beqa
I'_{3a1i}=\frac{\alpha_s^2}{\pi^2}  
\left\{\frac{1}{\epsilon^2}+\frac{1}{\epsilon} K_{\beta} \right\}
\eeqa
with counterterm
$I_{3a1i}^{c.t.}=(\alpha_s^2/\pi^2)  
[-2/\epsilon^2-K_{\beta}/\epsilon].$
Then
\beq
I_{3a1i}=\frac{\alpha_s^2}{\pi^2}  \frac{(-1)}{\epsilon^2}.
\label{I3a1i}
\eeq

Next we calculate $I'_{3a1ii}$. Using Eq. (\ref{A5}) we find 
\beq
I'_{3a1ii}=\frac{\alpha_s^2}{\pi^2}  
\left\{\frac{1}{2 \, \epsilon^2}+\frac{1}{2 \, \epsilon} \left[
1+K_{\beta}\right]\right\}
\eeq
with counterterm
$I_{3a1ii}^{c.t.}=(\alpha_s^2/\pi^2)  
[-1/\epsilon^2-K_{\beta}/(2 \epsilon)].$
Then
\beq
I_{3a1ii}=\frac{\alpha_s^2}{\pi^2}  
\left\{-\frac{1}{2 \, \epsilon^2}+\frac{1}{2 \, \epsilon}\right\}.
\label{I3a1ii}
\eeq
Adding Eqs. (\ref{I3a1i}) and (\ref{I3a1ii}) and denoting the sum by 
$I_{3a1}$, we find 
\beq
I_{3a1}=\frac{\alpha_s^2}{\pi^2} \left\{-\frac{3}{2\, \epsilon^2}
+\frac{1}{2\, \epsilon}\right\}.
\eeq

Last we calculate $I'_{3a2}$, given by
\beqa
&& \hspace{-3mm} 
I'_{3a2}=g_s^4 \int\frac{d^n k_1}{(2\pi)^n} \frac{d^n k_2}{(2\pi)^n}
\frac{v_i^{\mu}}{v_i\cdot k'} \, \frac{v_i^{\rho}}{v_i\cdot (k'-k_1)}  
\nonumber \\ && \times
\frac{v_i^{\nu}}{v_i\cdot (k'-k_1-k_2)} \, 
\frac{v_i^{\sigma}}{v_i\cdot (k'-k_2)}
\frac{(-i)g_{\mu \nu}}{k_1^2} \frac{(-i)g_{\rho \sigma}}{k_2^2} \, .
\nonumber \\ && 
\eeqa
Using Eqs. (\ref{A6}), (\ref{A7}), and (\ref{A8}), we find the UV poles of 
the integral 
\beq
I'_{3a2}=\frac{\alpha_s^2}{\pi^2} 
\left\{-\frac{1}{\epsilon^2}-\frac{1}{\epsilon} \left[\frac{1}{2}
+K_{\beta}\right]\right\}
\eeq
with counterterm
$I_{3a2}^{c.t.}=(\alpha_s^2/\pi^2)   
[2/\epsilon^2+K_{\beta}/\epsilon].$
Then
\beq
I_{3a2}=\frac{\alpha_s^2}{\pi^2}  
\left\{\frac{1}{\epsilon^2}-\frac{1}{2 \, \epsilon}\right\}.
\eeq

\subsection{$I_{3b}$}

In this section we calculate the three self-energy diagrams 
(with quark, gluon, and ghost loops) represented by 
Fig. 3(b). Note that an additional diagram with a four-gluon vertex vanishes.
 
After several manipulations the quark-loop diagram becomes
\beqa
&& \hspace{-4mm}
I'_{3bq}=-4 n_f  \frac{g_s^4}{(2\pi)^{2n}} \frac{1}{v_i \cdot k'}
\int \frac{d^n k \, d^n l}{k^4 \, v_i\cdot (k-k') \, l^2 \, (l-k)^2}
\nonumber \\ && \times 
\left[2(v_i\cdot l)^2-v_i^2 \, l^2 -2 v_i\cdot l \, v_i\cdot k 
+v_i^2 \, l\cdot k\right] \, .
\eeqa
Expanding around $v_i \cdot k'=0$ this gives 
\beqa
&& \hspace{-11mm}
I'_{3bq}=4 n_f \frac{g_s^4}{(2\pi)^{2n}} 
\int \frac{d^n k}{k^4 \, (v_i\cdot k)^2} 
\int \frac{d^n l}{l^2 \, (l-k)^2}
\nonumber \\ && \times 
\left[-v_i^2 \, l\cdot k-2(v_i\cdot l)^2+2 v_i\cdot l \, v_i\cdot k \right]\, .
\eeqa
A calculation of the UV poles of the integral using Eq. (\ref{A9}) gives 
\beq
I'_{3bq}=\frac{\alpha_s^2}{\pi^2} \frac{n_f}{3} 
\left\{-\frac{1}{\epsilon^2}-\frac{1}{\epsilon} \left[\frac{5}{6}
+K_{\beta}\right]\right\}
\eeq
with counterterm
$I_{3bq}^{c.t.}=(\alpha_s^2/\pi^2) (n_f/3) 
[2/\epsilon^2 + K_{\beta}/\epsilon]$.
Then
\beq
I_{3bq}=\frac{\alpha_s^2}{\pi^2} \frac{n_f}{3} 
\left[\frac{1}{\epsilon^2}-\frac{5}{6\, \epsilon}\right].
\eeq

Next we calculate the gluon-loop self-energy diagram 
in Fig. 3(b) which, 
after several manipulations, becomes
\beqa
&& \hspace{-10mm}
I'_{3bgl}=-\frac{g_s^4}{2(2\pi)^{2n}} \frac{1}{v_i \cdot k'}
\int\frac{d^n k}{v_i \cdot (k-k')}
\nonumber \\ && \times
\left\{\left[\frac{4 \, v_i^2}{k^2}
+(n-6)\frac{(v_i \cdot k)^2}{k^4}\right] \int \frac{d^n l}{l^2 (k-l)^2} \right.
\nonumber \\ && \quad \quad   
{}-(4n-6)\frac{v_i\cdot k}{k^4} \int d^n l \frac{v_i \cdot l}{l^2 (k-l)^2}
\nonumber \\ && \quad \quad \left. 
{}+\frac{(4n-6)}{k^4} 
\int d^n l \frac{(v_i \cdot l)^2}{l^2 (k-l)^2} \right\}.
\eeqa
Expanding around $v_i \cdot k'=0$ this gives 
\beqa
&& \hspace{-10mm}
I'_{3bgl}=-\frac{\alpha_s^2}{\pi^2} 2^{-5+2\epsilon} 
i \pi^{-2+\frac{3\epsilon}{2}} (1-\beta^2) 
\nonumber \\ && \hspace{-5mm} \times
\Gamma\left(\frac{\epsilon}{2}\right)
\left[\Gamma\left(1-\frac{\epsilon}{2}\right)\right]^2 
\frac{1}{\Gamma(2-\epsilon)} 
\nonumber \\ && \hspace{-5mm}
\times \left[2-\frac{(10-4\epsilon)}{8} \frac{1}{3-\epsilon}\right] 
\int \frac{d^n k}{(v_i \cdot k)^2 \, (k^2)^{1+\frac{\epsilon}{2}}}.
\eeqa
Using Eq. (\ref{A9}) the UV poles of the integral are then 
\beq
I'_{3bgl}=\frac{\alpha_s^2}{\pi^2} \frac{19}{48} 
\left\{-\frac{1}{\epsilon^2}-\frac{1}{\epsilon} \left[\frac{58}{57}
+K_{\beta}\right]\right\}.
\eeq
The counterterm is
$I_{3bgl}^{c.t.}=(\alpha_s^2/\pi^2) (19/48)  
[2/\epsilon^2 +K_{\beta}/\epsilon].$
Then
\beq
I_{3bgl}=\frac{\alpha_s^2}{\pi^2} \frac{19}{48} 
\left\{\frac{1}{\epsilon^2}-\frac{58}{57\, \epsilon}\right\}.
\label{I3bgl}
\eeq

Last we calculate the ghost-loop self-energy diagram 
in Fig. 3(b).
After several manipulations and 
expanding around $v_i \cdot k'=0$ this gives 
\beqa
&& \hspace{-8mm} 
I'_{3bgh}=\frac{\alpha_s^2}{\pi^2} 2^{-6+2\epsilon} 
\pi^{-2+\frac{3\epsilon}{2}} i (1-\beta^2) \Gamma\left(-1+\frac{\epsilon}{2}\right)
\nonumber \\ && \hspace{-3mm}\times 
\left[\Gamma\left(2-\frac{\epsilon}{2}\right)\right]^2 
\frac{1}{\Gamma(4-\epsilon)} 
\int \frac{d^n k}{(v_i \cdot k)^2 \, (k^2)^{1+\frac{\epsilon}{2}}}.
\eeqa
Using Eq. (\ref{A9}) the UV poles are then given by 
\beq
I'_{3bgh}=\frac{\alpha_s^2}{\pi^2} \frac{1}{48} 
\left\{-\frac{1}{\epsilon^2}-\frac{1}{\epsilon} \left[\frac{4}{3}
+K_{\beta}\right]\right\}
\eeq
with counterterm 
$I_{3bgh}^{c.t.}=(\alpha_s^2/\pi^2) (1/48)  
[2/\epsilon^2 +K_{\beta}/\epsilon].$
Then
\beq
I_{3bgh}=\frac{\alpha_s^2}{\pi^2} \frac{1}{48} 
\left\{\frac{1}{\epsilon^2}-\frac{4}{3 \, \epsilon}\right\}.
\label{I3bgh}
\eeq

The sum of the gluon and ghost loops, Eqs. (\ref{I3bgl}) and (\ref{I3bgh}), 
denoted by $I_{3bg}$, is then 
\beq
I_{3bg}=\frac{\alpha_s^2}{\pi^2} \frac{5}{12}
\left[\frac{1}{\epsilon^2}-\frac{31}{30\, \epsilon}\right].
\eeq

\subsection{$I_{3c}$}

Finally we calculate the diagram in Fig. 3(c). 
Using Eqs. (\ref{A1}) and (\ref{A2}) 
and adding the one-loop counterterm, we find 
\beq
I_{3c}= \frac{\alpha_s^2}{\pi^2} \frac{(1+\beta^2)}{2\beta} 
\frac{(-1)}{\epsilon^2}\ln\left(\frac{1-\beta}{1+\beta}\right).
\eeq

\section{Two-loop soft anomalous dimension}

We now combine the kinematic results from sections 3 and 4 
with color and symmetry factors. 
The contribution of the diagrams in Figs. (2) and (3) to the two-loop soft 
anomalous dimension is  
\beqa
&& C_F^2 \left[I_{2a}+I_{2b}+2 \, I_{2d}
+2 \, I_{2e}+I_{3a1}+I_{3a2}+I_{3c}\right]
\nonumber \\ && \hspace{-3mm}
{}+C_F \, C_A \left[-\frac{1}{2} I_{2b} 
+I_{2f} -I_{2cg}
-I_{2e}-I_{3bg}-\frac{1}{2} I_{3a2} \right]
\nonumber \\ && \hspace{-3mm}
{}+\frac{1}{2} C_F \left[I_{2cq}+I_{3bq}\right]  
\nonumber \\ &&\hspace{-3mm}
=\frac{\alpha_s^2}{\pi^2} \left[
-\frac{1}{2 \epsilon^2} \left(\Gamma_S^{(1)}\right)^2
+\frac{\beta_0}{4 \epsilon^2} \Gamma_S^{(1)}
-\frac{1}{2 \epsilon} \Gamma_S^{(2)} \right]. 
\label{S2}
\eeqa
On the right-hand side of Eq. (\ref{S2})  
in addition to $\Gamma_S^{(2)}$, which appears in the coefficient of the  
$1/\epsilon$ pole, there also  appear terms from the exponentiation 
of the one-loop result and the running of the coupling, 
with $\beta_0=(11/3) C_A-2n_f/3$, 
$C_A=N_c$,  which account for all the double poles of the graphs.
From Eq. (\ref{S2}) we solve for the two-loop soft anomalous dimension:
\beqa
&& \hspace{-5mm}\Gamma_S^{(2)}=\frac{K}{2} \, \Gamma_S^{(1)}
+C_F C_A \left\{\frac{1}{2}+\frac{\zeta_2}{2}
+\frac{1}{2} \ln^2\left(\frac{1-\beta}{1+\beta}\right) \right.
\nonumber \\ && \hspace{-5mm}
{}-\frac{(1+\beta^2)^2}{8 \beta^2} \left[\zeta_3
+\zeta_2 \ln\left(\frac{1-\beta}{1+\beta}\right)
+\frac{1}{3} \ln^3\left(\frac{1-\beta}{1+\beta}\right) \right.
\nonumber \\ && \hspace{-3mm}\left.
{}+\ln\left(\frac{1-\beta}{1+\beta}\right) 
{\rm Li}_2\left(\frac{(1-\beta)^2}{(1+\beta)^2}\right) 
-{\rm Li}_3\left(\frac{(1-\beta)^2}{(1+\beta)^2}\right)\right] 
\nonumber \\ && \hspace{-5mm} 
{}-\frac{(1+\beta^2)}{4 \beta} \left[\zeta_2
-\zeta_2 \ln\left(\frac{1-\beta}{1+\beta}\right) 
+\ln^2\left(\frac{1-\beta}{1+\beta}\right)\right.
\nonumber \\ && 
{}-\frac{1}{3} \ln^3\left(\frac{1-\beta}{1+\beta}\right)
+2  \ln\left(\frac{1-\beta}{1+\beta}\right) 
\ln\left(\frac{(1+\beta)^2}{4 \beta}\right) 
\nonumber \\ &&  \hspace{15mm} \left. \left.
{}-{\rm Li}_2\left(\frac{(1-\beta)^2}{(1+\beta)^2}\right)\right]\right\}
\label{Gamma2}
\eeqa
where
\beq
K=C_A\left(\frac{67}{18}-\zeta_2 \right)-\frac{5n_f}{9}.
\eeq
A slightly different but equivalent expression is given in Ref. \cite{NKprl}.

In terms of the cusp angle $\gamma$ we find 
\beqa
&& \hspace{-5mm} \Gamma_S^{(2)}=\frac{K}{2} \, \Gamma_S^{(1)}
+C_F C_A \left\{\frac{1}{2}+\frac{\zeta_2}{2}+\frac{\gamma^2}{2} 
-\frac{1}{2}\coth^2\gamma \right.
\nonumber \\ && \quad \times
\left[\zeta_3-\zeta_2\gamma-\frac{\gamma^3}{3}
-\gamma \, {\rm Li}_2\left(e^{-2\gamma}\right)
-{\rm Li}_3\left(e^{-2\gamma}\right)\right] 
\nonumber \\ && \hspace{-4mm}
{}-\frac{1}{2} \coth\gamma\left[\zeta_2+\zeta_2\gamma+\gamma^2
+\frac{\gamma^3}{3} \right.
\nonumber \\ && \hspace{14mm} \left. \left. 
{}+2\, \gamma \, \ln\left(1-e^{-2\gamma}\right)
-{\rm Li}_2\left(e^{-2\gamma}\right)\right] \right\}.
\eeqa
This expression is consistent with the form of the two-loop cusp 
anomalous dimension given in Ref. \cite{KR} (which involves a few 
uncalculated integrals), 
and it is also consistent with the two-loop heavy-quark form factor 
of Ref. \cite{HQFF}. 

Using the above results we  can find the soft-gluon logarithms in the cross 
section for $e^+ e^- \rightarrow t {\bar t}$, which are  
of the form $\ln^{n-1}(\beta^2)/\beta^2$ at $n$-th order in $\alpha_s$. 
The first-order soft-gluon corrections are 
$\sigma^{(1)}=\sigma^B \, (\alpha_s/\pi) \, 2 \, \Gamma_S^{(1)} / \beta^2$
with $\sigma^B$ the Born cross section.
The second-order soft-gluon corrections are
\beqa
&& \hspace{-8mm}
\sigma^{(2)}=\sigma^B \frac{\alpha_s^2}{\pi^2} 
\left\{ \left[4 (\Gamma_S^{(1)})^2-\beta_0  \,\Gamma_S^{(1)} \right] 
\frac{\ln(\beta^2)}{\beta^2} \right.
\nonumber \\ && \hspace{15mm} \left.
{}+\left[2 \,T_1 \, \Gamma_S^{(1)} +2 \, \Gamma_S^{(2)}\right]
\frac{1}{\beta^2} \right\}
\eeqa
with $T_1$ the NLO virtual corrections.

\begin{figure}
\begin{center}
\includegraphics[width=8cm]{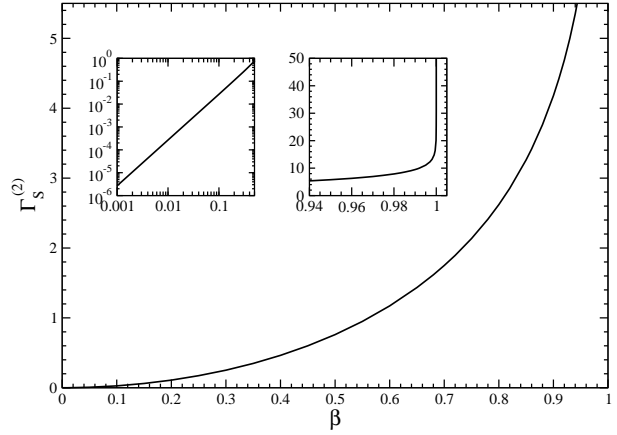}
\caption{The two-loop soft anomalous dimension $\Gamma_S^{(2)}$.}
\label{Gamma2Kexp}
\end{center}
\end{figure}

In Figure 5 we plot $\Gamma_S^{(2)}$ versus $\beta$. The insets 
show the small and large $\beta$ regions in detail.
Note that both $\Gamma_S^{(1)}$ and $\Gamma_S^{(2)}$ vanish in 
the threshold limit, $\beta=0$, 
and they both diverge in the massless limit, $\beta=1$.

The small-$\beta$ expansion of Eq. (\ref{Gamma2}) is 
\beqa
\Gamma_{S \, \rm exp}^{(2)}&=&-\frac{2}{27} \beta^2 
\left[C_F C_A (18 \zeta_2-47)+5 n_f C_F\right]
\nonumber \\ && \hspace{30mm}
{}+{\cal O}(\beta^4) \, .
\eeqa
$\Gamma_S^{(2)}$ is an even function of $\beta$ and hence 
only even powers of $\beta$ appear in the expansion.
As shown in \cite{NKprl} the expansion 
provides a good approximation to the complete result for small $\beta$.

Next we study the large-$\beta$ behavior of $\Gamma_S^{(2)}$, 
Eq. (\ref{Gamma2}). 
As $\beta \rightarrow 1$, 
\beq 
\Gamma_S^{(2)} \rightarrow \frac{K}{2} \Gamma_S^{(1)}
+C_F C_A \frac{(1-\zeta_3)}{2}
\eeq
with $\Gamma_S^{(1)}=C_F\left[\ln\left(2 v_i \cdot v_j/\sqrt{v_i^2 v_j^2}
\right)-1\right]$. Since $\Gamma_S^{(1)}$ diverges at $\beta=1$,  
$\Gamma_S^{(2)} \approx (K/2) \Gamma_S^{(1)}$ at that 
limit. This is consistent  with the massless case $\Gamma_S^{(2)}=\frac{K}{2} \Gamma_S^{(1)}$ 
with $\Gamma_S^{(1)}=C_F \ln(v_i \cdot v_j)$ \cite{ADS,BN,GM}.
As is clear from Eq. (\ref{Gamma2}) the massive case is much more complicated
than  the simple massless relation, and $(K/2) \Gamma_S^{(1)}$ is just the 
first of many terms in the expression for $\Gamma_S^{(2)}$. 
Figure 6 shows that numerically the ratio  
$(K/2) \Gamma_S^{(1)}/\Gamma_S^{(2)}$ goes to 
1 as $\beta \rightarrow 1$, the massless limit, as expected from the above 
discussion, but it is significantly different from 1 at other $\beta$  
and takes the value 1.144 at the other end of the range, $\beta=0$. 

In the mixed massive-massless case,  with $v_i$ the heavy quark and $v_j$ the 
massless quark, we find 
\beq
\Gamma_S^{(2)}=\frac{K}{2} \Gamma_S^{(1)}
+C_F C_A \frac{(1-\zeta_3)}{4} 
\eeq
with 
\beq
\Gamma_S^{(1)}=C_F \left[\ln\left(\frac{\sqrt{2} \, v_i \cdot v_j}
{\sqrt{v_i^2}}
\right)-\frac{1}{2}\right].
\eeq 

Given that the small-$\beta$ expansion gives very good 
approximations to $\Gamma_S^{(2)}$ at smaller $\beta$ while the 
expression $(K/2)\, \Gamma_S^{(1)}$ is the large-$\beta$ limit, 
we can derive an approximation to $\Gamma_S^{(2)}$, Eq. (\ref{Gamma2}), 
valid for all $\beta$ using the following approximate formula:
\beqa
&& \hspace{-10mm} \Gamma^{(2)}_{S \, \rm approx}=\Gamma^{(2)}_{S \, \rm exp}
+\frac{K}{2} \Gamma_S^{(1)}-\frac{K}{2} \Gamma^{(1)}_{S \, \rm exp}
\nonumber \\ && \hspace{-3mm}
=\frac{K}{2} \Gamma_S^{(1)}+C_F C_A \left(1-\frac{2}{3}\zeta_2\right) \beta^2
+{\cal O}\left(\beta^4\right).
\eeqa

\begin{figure}
\begin{center}
\includegraphics[width=8cm]{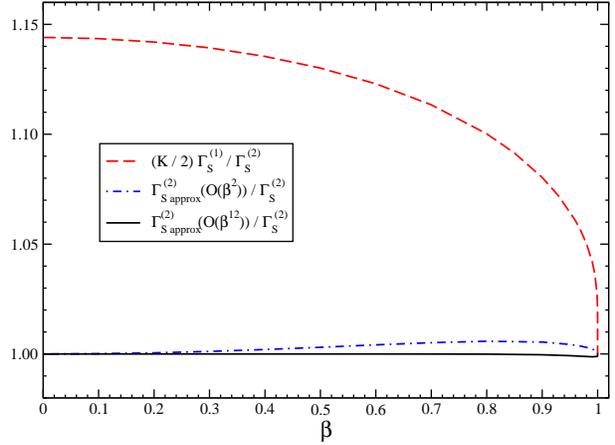}
\caption{Approximations to $\Gamma_S^{(2)}$.}
\label{Gamma2approx}
\end{center}
\end{figure}

As seen from Fig. 6, adding just the $\beta^2$ terms to 
$(K/2)\, \Gamma_S^{(1)}$  
provides an excellent approximation to the exact result for 
for all $\beta$. 
If we keep additional terms through order $\beta^{12}$ then the approximation 
is extremely good and does not differ by more than one per mille anywhere 
in the $\beta$ range.

Finally we discuss extensions to single top \cite{NKtop,NKst} and 
top pair \cite{NKtop,NKRV} production at hadron colliders.  
As first shown in Ref. \cite{NKprl}, 
since the two-loop soft anomalous 
dimensions for single top and top pair processes involve eikonal graphs with 
both massless and massive eikonal lines,  
it is clear that the simple massless relation  
$\Gamma_S^{(2)}=(K/2)\, \Gamma_S^{(1)}$ will not hold.
This observation was also made later in Refs. \cite{MSS,BNm}. 
The two-loop soft anomalous dimension $\Gamma_S^{(2)}$ 
derived in \cite{NKprl} and in this paper is an essential ingredient 
for next-to-next-to-leading logarithm (NNLL) resummation for 
heavy quark hadroproduction   
and has been used in very recent calculations \cite{BNm,BFS,CMS,FNPY}.

\begin{acknowledgments}
This work was supported by the National Science Foundation under 
Grant No. PHY 0855421.
\end{acknowledgments}

\section*{Appendix}

We list results for the UV poles of several of the many integrals needed 
in the calculation of the two-loop soft anomalous dimension.
 
\beqa
&& \hspace{-7mm}
\int\frac{d^n k}{k^2 \, v_i\cdot k \, v_j\cdot k}=\frac{i}{\epsilon}
(-1)^{-1-\frac{\epsilon}{2}} \pi^{2-\frac{\epsilon}{2}} 
2^{3+3\frac{\epsilon}{2}}
\nonumber\\ && \quad \times
\Gamma\left(1+\frac{\epsilon}{2}\right)\, {}_2F_1\left(\frac{1}{2},
1+\frac{\epsilon}{2};\frac{3}{2};\beta^2\right)
\nonumber\\ &&
=\frac{i 4\pi^2}{\beta}
\left\{\frac{1}{\epsilon}\ln\left(\frac{1-\beta}{1+\beta}\right) \right.
\nonumber\\ && \quad
{}+\frac{1}{2}(2\ln 2-\ln \pi-\gamma_E-i \pi) 
\ln\left(\frac{1-\beta}{1+\beta}\right) 
\nonumber\\ && \quad 
{}+\frac{1}{4}\ln^2(1+\beta)-\frac{1}{4}\ln^2(1-\beta)
\nonumber\\ && \quad \left. 
{}-\frac{1}{2}{\rm Li}_2\left(\frac{1+\beta}{2}\right)
+\frac{1}{2}{\rm Li}_2\left(\frac{1-\beta}{2}\right) \right\} 
+{\cal O}(\epsilon)
\label{A1}
\nonumber \\
\eeqa
where ${}_2F_1$ is the Gauss hypergeometric function.

\beqa
&& \hspace{-9mm}
\int\frac{d^n k}{k^2 \, (v_i\cdot k)^2}
=\frac{i}{\epsilon}\,
(-1)^{1-\frac{\epsilon}{2}} \, \pi^{2-\frac{\epsilon}{2}} \,  
2^{3+3\frac{\epsilon}{2}} 
\nonumber\\ && \quad \quad \times 
(1-\beta^2)^{-1-\frac{\epsilon}{2}} 
\Gamma\left(1+\frac{\epsilon}{2}\right)
\nonumber\\ && \hspace{-8mm}
=-\frac{i 8 \pi^2}{1-\beta^2}
\left[\frac{1}{\epsilon} -\frac{1}{2}\ln(1-\beta^2)\right.
\nonumber\\ && \hspace{7mm} \left. 
{}+\frac{3}{2} \ln 2-\frac{1}{2}\ln \pi-\frac{\gamma_E}{2}
-\frac{i \pi}{2} \right] 
+{\cal O}(\epsilon).
\label{A2}
\eeqa

\beqa
&& \hspace{-9mm}
\int\frac{d^n k}{(k^2)^{1+\frac{\epsilon}{2}} \, v_i\cdot k \, v_j\cdot k}
=\frac{i}{\epsilon^2} \, \frac{(-1)^{1-\epsilon}}{\beta} \, 2^{2\epsilon} 
\nonumber\\ && \hspace{-4mm} \times
\pi^{2-\frac{\epsilon}{2}} \, \Gamma(1+\epsilon) 
\frac{1}{\Gamma\left(1+\frac{\epsilon}{2}\right)} 
\nonumber\\ &&  \hspace{-4mm} \times 
\left[(1-\beta)^{-\epsilon} {}_2F_1\left(-\epsilon,
1+\epsilon;1-\epsilon;\frac{1-\beta}{2}\right) \right.
\nonumber\\ &&  \hspace{-2mm} \left. 
{}-(1+\beta)^{-\epsilon} {}_2F_1\left(-\epsilon,
1+\epsilon;1-\epsilon;\frac{1+\beta}{2}\right)\right].
\label{A3}
\eeqa

\beqa
&& \hspace{-9mm}
\int\frac{d^n k}{k^2 \, (v_i\cdot k)^{1+\epsilon} \, v_j\cdot k}
=\frac{i \pi^{2-\frac{\epsilon}{2}}}{\epsilon (1+\epsilon)} 
2^{2+\frac{9\epsilon}{2}} (-1)^{-1-\frac{3\epsilon}{2}}
\nonumber\\ && \hspace{-6mm} \times 
(1-\beta^2)^{-1-\frac{3\epsilon}{2}} \, 
\Gamma\left(1+\frac{3\epsilon}{2}\right) \frac{1}{\Gamma(1+\epsilon)}
\nonumber\\ && \hspace{-6mm} \times 
F_1[1+\epsilon;1+\frac{3\epsilon}{2},1+\frac{3\epsilon}{2};2+\epsilon;
\frac{2\beta}{1+\beta},\frac{-2\beta}{1-\beta}] 
\label{A4}
\eeqa
where $F_1$ is the Appell hypergeometric function.

\beqa
&& \hspace{-7mm}
\int\frac{d^n k_2}{k_2^2 \, \left[v_i\cdot (k_1+k_2)\right]^2}
=\frac{i}{\epsilon } \, \frac{(-1)^{-1+\frac{\epsilon}{2}}}{(1+\epsilon)} \, 
2^{4-\frac{\epsilon}{2}} \, \pi^{\frac{3-\epsilon}{2}} 
\nonumber\\ && \quad \times 
(1-\beta^2)^{-1+\frac{\epsilon}{2}} \, (v_i \cdot k_1)^{-\epsilon} \, 
\Gamma\left(1+\frac{\epsilon}{2}\right) 
\nonumber\\ && \quad \times 
\Gamma\left(1-\frac{\epsilon}{2}\right) \, 
\Gamma\left(\frac{3+\epsilon}{2}\right)  \, .
\label{A5}
\eeqa

\beqa
&& \hspace{-7mm}
\int\frac{d^n k}{k^2 \, (v_i\cdot k)^{2+\epsilon}}
=\frac{i \pi^{2-\frac{\epsilon}{2}}}{\epsilon (1+\epsilon)} 
2^{2+\frac{9\epsilon}{2}} (-1)^{-1-\frac{3\epsilon}{2}}
\nonumber\\ && \quad \times 
(1-\beta^2)^{-1-\frac{3\epsilon}{2}} \, 
\Gamma\left(1+\frac{3\epsilon}{2}\right) \frac{1}{\Gamma(1+\epsilon)} \, .
\label{A6}
\eeqa

\beqa
&& \hspace{-7mm}
\int\frac{d^n k_1}{k_1^2 \, v_i\cdot k_1 \, v_i\cdot (k_1+k_2)}
=\frac{i}{\epsilon} \, (-1)^{\frac{\epsilon}{2}} \, 
2^{2-\frac{\epsilon}{2}} \, \pi^{\frac{3-\epsilon}{2}} 
\nonumber\\ && \quad \quad \times 
(v_i \cdot k_2)^{-\epsilon} \, (1-\beta^2)^{-1+\frac{\epsilon}{2}} \,  
\Gamma\left(1+\frac{\epsilon}{2}\right) 
\nonumber\\ && \quad \quad \times
\Gamma\left(1-\frac{\epsilon}{2}\right) \, 
\Gamma\left(\frac{\epsilon-1}{2}\right)  \, .
\label{A7}
\eeqa

\beqa
&& \hspace{-7mm}
\int\frac{d^n k_2}{k_2^2 \, v_i\cdot k_2 \, [v_i\cdot (k_1+k_2)]^2}
=\frac{i}{1-\epsilon} \, (-1)^{1+\frac{\epsilon}{2}} 
\nonumber\\ && \quad \times 
2^{3-\frac{3\epsilon}{2}} \, \pi^{2-\frac{\epsilon}{2}} \, 
(v_i \cdot k_1)^{-1-\epsilon} \, (1-\beta^2)^{-1+\frac{\epsilon}{2}} 
\nonumber\\ && \quad \times 
\Gamma\left(1-\frac{\epsilon}{2}\right) \, \Gamma(1+\epsilon)  \, .
\label{A8}
\eeqa

\beqa
&& \hspace{-7mm}
\int\frac{d^n k}{(k^2)^{1+\frac{\epsilon}{2}} \, (v_i\cdot k)^2}
=\frac{i}{\epsilon} \, (-1)^{-1-\epsilon} \, 
2^{2+3\epsilon} \, \pi^{2-\frac{\epsilon}{2}} 
\nonumber\\ && \quad \quad \times 
(1-\beta^2)^{-1-\epsilon} \,  \Gamma(1+\epsilon) \, 
\frac{1}{\Gamma\left(1+\frac{\epsilon}{2}\right)}  \, .
\label{A9}
\eeqa

\end{document}